\documentclass[aps,prb,twocolumn,showpacs,floatfix]{revtex4}
\usepackage{graphicx}

\begin{document}

\title{An extended Falicov-Kimball model on a triangular lattice}
\author{Umesh K. Yadav, T. Maitra, Ishwar Singh}
\affiliation{Department of Physics, Indian Institute of Technology Roorkee, Roorkee- 247667, Uttarakhand, India}
\author{A. Taraphder}
\affiliation{Department of Physics and Centre for Theoretical Studies, Indian Institute of Technology, Kharagpur - 721 302, India}
\pacs {}
\date{\today}

\begin{abstract}
The combined effect of frustration and correlation in electrons
is a matter of considerable interest of late. In this context
a Falicov-Kimball model on a triangular lattice with two localized states,
relevant for certain correlated systems, is considered.
Making use of the local symmetries of the model, our numerical
study reveals a number of orbital ordered ground states, tuned by the small
changes in parameters while quantum fluctuations between the localized and
extended states produce homogeneous mixed valence. The inversion symmetry
of the Hamiltonian is broken by most of these ordered states leading to
orbitally driven ferroelectricity. We demonstrate that
there is no spontaneous symmetry breaking when the ground state is
inhomogeneous. The study could be relevant for frustrated systems like
$GdI_2$, $NaTiO_2$ (in its low temperature C2/m phase) where two Mott
localized states couple to a conduction band.
\end{abstract}

\vspace{0.5cm}
\maketitle
\section{Introduction}

Geometric frustration in correlated systems brings about a variety of
phenomena and is a major area of interest in the condensed matter
community presently. Systems having two dimensional (2D) layered structure with
triangular lattice (for example, transition metal dichalcogenides~\cite{aebi,qian2,cava},
cobaltates \cite{qian}, GdI$_2$ \cite{gd1,gd2}and its doped variant~\cite{gd3,simon},
NaTiO$_2$~\cite{clark,khom}, NaVO$_2$ \cite{jia} etc.) are known to
come up with a host of cooperative phenomena like valence and metal
insulator transitions, charge, orbital and magnetic order, unconventional
superconductivity, excitonic instability \cite{aebi} and possible non-Fermi
liquid states\cite{gd2,castro}. These systems pose a challenge to theoretical
understanding as the underlying geometric frustration of the triangular lattice,
coupled with strong dynamic fluctuations, give rise to a large degeneracy at
low temperatures and competing ground states close by in energy. A consequence
of this is a fairly complex phase diagram \cite{gd3} and the presence of soft
local modes strongly coupled to the itinerant electrons \cite{gd2}.

\begin{figure*}[ht]
\begin{center}
\vspace{-1.60cm}
\includegraphics[width=12.10cm]{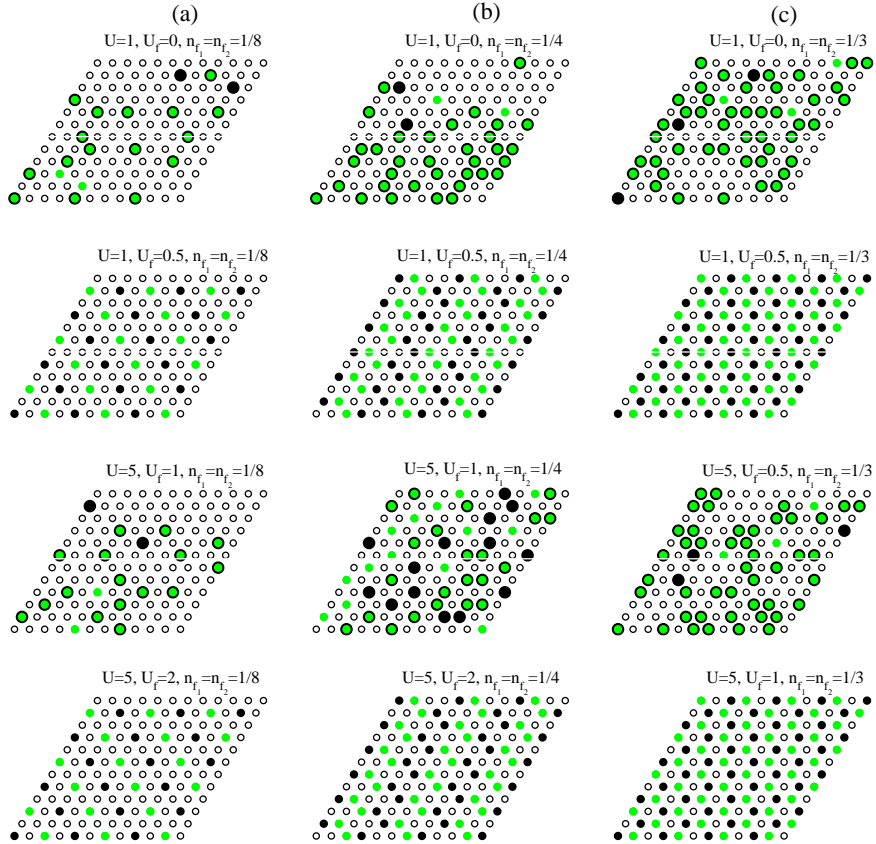}
\caption{ (colour online) $f$-electron configurations are shown at each site
for (a) $n_{f_{1}}$ = $n_{f_{2}}$ = 1/8, (b) $n_{f_{1}} = n_{f_{2}} = 1/4$ and (c) $n_{f_{1}} = n_{f_{2}} = 1/3$.
Black and green circles correspond to sites occupied by $f_1$ and $f_2$-electrons respectively
and open circles correspond to sites with no $f$-electron occupancy.}
\end{center}
\end{figure*}

We motivate the model by looking at two systems GdI$_2$ and NaTiO$_2$.
Each layer of Gd ions in GdI$_2$ form a 2D triangular lattice. They
are well separated from each other by intervening layers of large Iodine
ions and do not interact significantly with each other. From
recent band structure (LSDA) calculations\cite{gd1,gd3} it is known
that three nearly degenerate, spin-polarized $d$-orbitals (d$_{z^2}$, d$_{x^2-y^2}$ and
d$_{xy}$) cross the Fermi level. Calculations involving dynamical local
correlations\cite{gd2} show that these three nearly degenerate $d$-orbitals
further break down to two doubly degenerate (d$_{x^2-y^2}$ and d$_{xy}$)
localized levels below and one extended (d$_{z^2}$) level across the Fermi
level. From experimental studies~\cite{simon}, it is observed that the ground state
is insulating and is likely to be orbitally ordered in a three sublattice
fashion at half filling (each orbital occupying one sublattice) while a small doping away
from half-filling leads to phase segregation. An effective Falicov-Kimball model~\cite{fkm}
of spinless Fermions with two (degenerate) localized bands and one itinerant band
is proposed recently for this system~\cite{gd2}. It would, therefore, be interesting
to look for such orbitally ordered states in this context.

The corresponding Hamiltonian, then, is

\begin{eqnarray}
{H} =-\sum_{\langle ij\rangle}(t_{ij}+\mu\delta_{ij})d^{\dagger}_{i}d_{j}
+E_f \sum_{i,\alpha=1,2} (f^{\dagger}_{i\alpha}f_{i\alpha})
\nonumber \\
+ U \sum_{i,\alpha=1,2}{(f^{\dagger}_{i\alpha}f_{i\alpha}d^{\dagger}_{i}d_{i})}
+ U_f \sum_{i}{(f^{\dagger}_{i1}f_{i1}f^{\dagger}_{i2}f_{i2})}.
\end{eqnarray}

\noindent Here $d^{\dagger}_{i}, d_{i}$  are, respectively, the creation and
annihilation operators for electrons in the itinerant band
and $f{\dagger}_{i\alpha}, f_{i\alpha}$ are the same for the two
localized bands at the site $i$. The first term in Eq.(1) is the kinetic
energy of $d$-electrons on a triangular lattice (only nearest-neighbor
hopping is considered) while the second term represents the degenerate energy
levels $E_{f}$ of the $f_{1}$ and $f_{2}$ electrons. The third term is
the on-site Coulomb repulsion between $d$- and $f$-electrons. The last one
is the local repulsion between the $f$-electrons.

\begin{figure}
\begin{center}
\vspace{0.63cm}
\includegraphics[width=7.5cm,height=5.9cm]{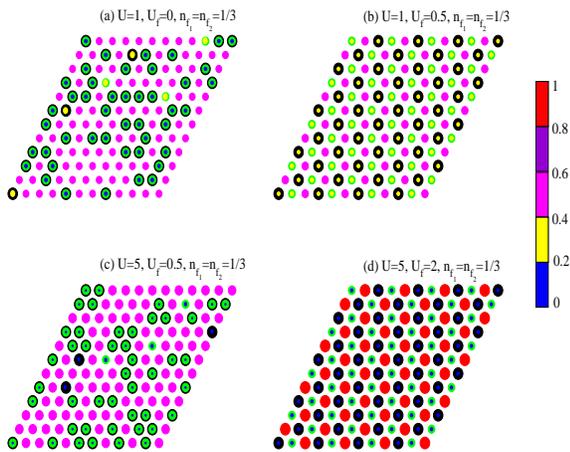}
\caption{ (colour online) d-electron densities for different $U$ and $U_f$ values (set of four
figures (a) to (d)) for $n_{f_{1}}$ = $n_{f_{2}}$ = 1/3.}
\end{center}
\end{figure}

NaTiO$_2$ is another system where the above model can be applied. It
also has a layered structure with alternating NaO and TiO slabs and shows
a correlation-driven metal-insulator transition~\cite{khom}.
The Ti$^{3+}$ ions with one electron in the $d$-orbitals are arranged on a
triangular lattice and are octahedrally coordinated to oxygen ions. In
the low temperature monoclinic phase the TiO$_6$ octahedron gets distorted,
leading to a splitting in the triply degenerate t$_{2g}$ orbitals of Ti 3$d$:
two orbitals are pushed below the Fermi energy while one straddles it.
Orbital order due to electronic correlations has been predicted in
this system earlier~\cite{khom}. While the absence of magnetic order renders a
consideration of spin degeneracy redundant, correlations in the 3$d$ band would
further localize the two bands below Fermi level and the effective model above
could as well describe the low energy dynamics of this system.

The Hamiltonian conserves local $f$-electron occupation
numbers $\hat{n}_{f,i\alpha}=f^{\dagger}_{i\alpha}f_{i\alpha}$ owing to the
local $U(1)$ gauge invariance in the absence of $f$-$d$ hybridization.
Therefore, $[\hat{n}_{f,i\alpha},H]=0$ and $\omega_{i\alpha}=
f^{\dagger}_{i\alpha}f_{i\alpha}$ are good quantum numbers taking values
only 0 or 1. The local conservation also implies that the Hamiltonian may be
written as,
\begin{equation}
H=\sum_{\langle ij\rangle}h_{ij}(\omega)d^{+}_{i}d_{j}+E_f\sum_{i,\alpha}
\omega_{i,\alpha} + U_{f}\sum_{i}\omega_{i1}\omega_{i2}
\end{equation}
\noindent where $h_{ij}(\omega)=-t_{ij}+(U\sum_{\alpha}\omega_{i,\alpha}-\mu)
\delta_{ij}$.

We set the scale of energy as the nearest neighbor hopping $t=1$. The eigenvalue
spectrum of this Hamiltonian, is easily obtained by numerical diagonalization on
a finite size triangular lattice with periodic boundary condition. In order to calculate
the average values of physical quantities, the classical Monte Carlo method
using Metropolis algorithm can then be employed by `annealing' over a subset of
configurations of the ``classical" variables
$\{\omega_{i,\alpha}\}$. This approach is known to give reliable results
for the FKM, even on a triangular lattice with macroscopic degeneracies.
The details of the method is reported elsewhere\cite{uky}.

\begin{figure}
\begin{center}
\vspace{0.0cm}
\includegraphics[width=7.0cm]{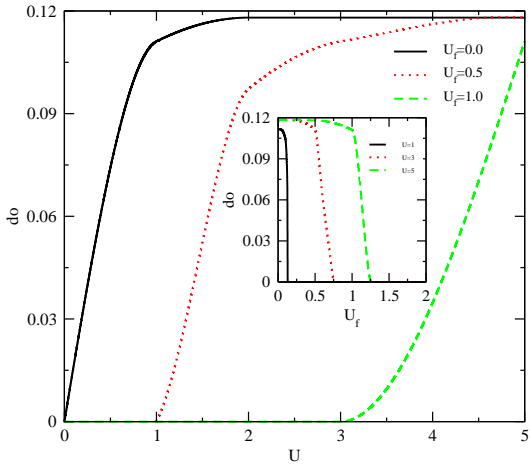}
\caption{ (colour online) Number of sites doubly occupied by localized f-electrons versus $U$ for different $U_f$
(versus $U_f$ for different $U$ (inset)) for $n_{f_{1}}$ = $n_{f_{2}}$ = 1/8.}
\end{center}
\end{figure}

For the systems discussed above, the number of electrons in the $d$-band is
one at every site in the undoped limit. Therefore, in the following
calculations, we restrict ourselves to the
half-filled limit, i.e. $N_{f} + N_{d} = N$ where $N_{f},\, N_{d}$ and $N$
are the total number of $f$-electrons, $d$-electrons and the number
of sites in the lattice respectively (we have used $N=144$ in all
the calculations and checked a few results with larger $N$). In addition,
$N_{f1} = N_{f2} = N_{f}/2$, as the two $f$-levels are degenerate.

\begin{figure*}[ht]
\begin{center}
\vspace{-0.450cm}
\includegraphics[width=11.5cm]{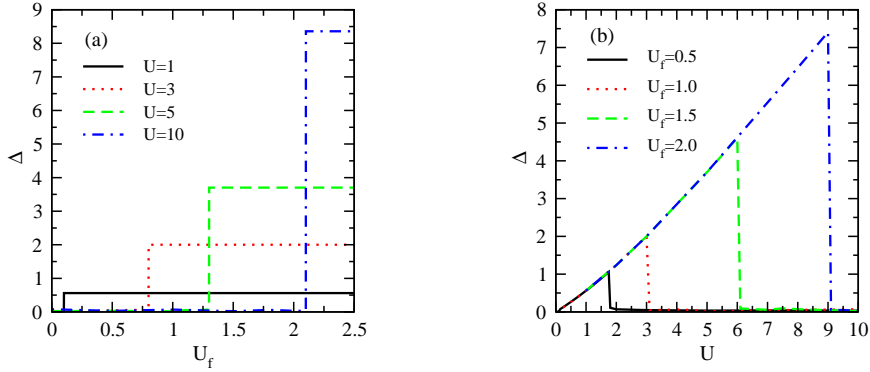}
\caption{(colour online) Energy gap $\Delta$ (a) as a function of $U_f$ for different values of $U$ (b) as a
function of $U$ for different $U_f$ at $n_{f_{1}}$ = $n_{f_{2}}$ = 1/8.}
\end{center}
\end{figure*}

\section{Results and discussion}

In Fig.1 we show the ground state configurations for different $U$ and $U_f$
at various $f$ electron concentrations. Each column shows the progression  
of states from disorder to order to disorder and finally order again as a 
function of $U$ and $U_f$ for different $n_f$ values. Fig.1(a) presents the case
for $n_{f_{1}}\,(=\frac{N_{f_{1}}}{N})=n_{f_{2}}=1/8$. In the absence of any
interaction between the two
localized $f$-electron states ($U_f=0$), no particular order is observed
even on increasing the value of $U$. In fact, a finite $U$ favors double
occupancies of $f$-electrons, leaving a larger part of the lattice for
$d$-electron motion. Therefore, with increasing $U$ there is an increasing tendency of
phase separation in real space between localized and itinerant
electrons. Keeping $U$ small (for example, $U=1$) and making  $U_f$ finite
begins to remove the local double occupancies of $f$ electrons, reduces the
overall kinetic energy of $d$-electrons and leads to an ordered stripe-like
pattern of the two localized electrons ($f_1$ and $f_2$). This stripe
pattern appears at a higher $U_f$ when $U$ is raised. A transition
from a state of disordered doubly occupied sites to a stripe-like ordered
pattern as a function of $U_f$ at fixed $U$ is shown in Fig.1.
This is a discontinuous transition appearing at a critical value of $U_f$
(which increases with $U$) and the ordered phase remains stable up to
high values of $U_f$. Note that on raising $U_f$, the double occupancy
is removed very quickly (see later) and the ground state is fairly insensitive
to a rise in $U_f$ further. Similar trends are seen for other fillings too
(e.g., $n_{f_{1}}=n_{f_{2}}=1/4$ and $n_{f_{1}}=n_{f_{2}}=1/3$, shown in Fig.1(b) and
1(c) respectively), albeit with different real space patterns of the $f$
electrons: for $n_{f_{1}}=n_{f_{2}}=1/4$ it is a bi-stripe orbital order whereas for
$n_{f_{1}}=n_{f_{2}}=1/3$ it is a three sub-lattice structure involving $f_1$, $f_2$
and empty (no $f$-electron) sites decorating the vertices of each triangle.

One of the most interesting consequences of the various orbital orders seen
in Fig.1 is that many of them break the inversion symmetry of the Hamiltonian.
This will induce spontaneous displacements of the ions and the resulting lattice
distortion will lead to ferroelectricity with a spontaneous finite polarisation. 
This is due to a specific realization of the unconventional orbital order 
that breaks inversion symmetry. It is driven by electronic correlation, very different 
from the conventional route to ferroelectricity.

In Fig.2 we present the d-electron densities at each site. It is interesting to
observe how the competition among different interaction energies (such as
kinetic energy of the d-electrons, on-site correlation energies between d- and
f-electrons as well as between two localized f-electrons) affect the transport
of the system. The $d$-electron density plots at each site (presented in Fig.2
for $n_{f_{1}} = n_{f_{2}} = 1/3$) clearly show the appearance of metallic
and insulating and re-entrant metallic phases by
tuning the parameters $U$ and $U_f$. From Figs.2; (b) and (c)  we see
that on increasing $U$ to a high value, keeping $U_f$ fixed, the system
goes from a stripe like ordered configuration to a phase separated state.
This transition is accompanied as well by an insulator to metal transition
as we discuss below. Again by keeping $U$ fixed and increasing $U_{f}$, the
ordered structure reappears and the system becomes insulating (Figs.2; (c),
(d)). By tuning the interaction parameters $U$
and $U_f$ one can switch between these metallic and insulating states.
Note that in some cases there are also sites substantially occupied by electrons 
of all three types ($f_{1}, \, f_ {2}$ and $d$). Number of sites doubly occupied by 
localized f-electrons predominantly plays a crucialrole in determining 
the critical $U$ for the
metal-insulator transitions discussed above. We show the dependence of this
double occupancy ($do$) on the interaction parameters $U$ and $U_f$ for $n_{f1} =
n_{f2} = 1/8$ in Fig.3. The double occupancy $do$
is extremely sensitive to $U_f$, dropping off to zero rapidly for $U_f$ as small
as 0.25 for various value of $U$ and $n_{f,\alpha}$. It increases
with $U$ as expected and saturates quite rapidly thereafter. These features are 
quite general and are obsereved at other fillings of $n_{f,\alpha}$ also.  

In order to investigate the electronic properties of these ordered phases
further we have calculated the gap~\cite{gap_defn} in the spectrum for each of
these phases. Fig.4(a) shows the gap as a function of $U_f$ for a
series of $U$ values, for the filling $1/8$. The disorder to order transition
is indeed accompanied by a metal to insulator transition (MIT), the disordered
phase has no gap at the
the Fermi energy, whereas the ordered states are insulating with a finite gap.
In addition, the gap appears at a critical $U_f$ and remains constant beyond
(independent of $U_f$ as long as $U$ is fixed). The discontinuous nature of
the transition is clearly visible in Fig.4.
On increasing $U$ the same feature is observed, though the critical $U_f$
for MIT increases. As $U_f$ is the correlation energy between two localized
$f$-electrons and $U$ is the same between localized and itinerant electrons,
there is a competition between these two interactions and therefore the 
critical $U_f$ depends strongly on $U$. In Figs. 4(a) we show the energy 
gap as a function of $U_f$ at fixed values of $U$ for $n_{f1} = n_{f2} = 1/8$.
An interesting observation can be gleaned from the energy gaps if we keep $U_f$ 
fixed (say $U_f=1$) and vary $U$. For $U = 0$ there is no gap in the spectrum 
and on increasing $U$, the
gap increases almost linearly with $U$ (Fig.4(b)). Above a certain value of $U$
(for $U_{f}=1$, at around $U=3.2$), however, the gap suddenly drops to zero
and the system becomes metallic. Therefore, as a function of $U$ and at a
fixed $U_f$, there is a metallic phase appearing at large $U$. This seemingly
counter-intuitive result is understood from the fact that there are two
competing processes at hand (coming from $U$ and $U_f$). At $U=0$ the $d$-electrons
are free and the system is metallic. As $U$ increases they avoid sites
occupied by $f$-electrons and a gap appears. There is also a tendency
towards $f$-$d$ phase segregation as discussed above. A finite value of $U_f$, however,
strongly disfavors double occupancy (as in Fig.3) and spreads the $f$-electrons out.
Therefore, when, at a higher $U$, double occupancy (with $f_1$ and $f_2$ electrons)
is energetically favorable over a joint occupation by $f_{\alpha}$ and $d$ electrons
at a site, the metallic state reappears.  
We have also calculated the corresponding density of states for different
values of $U$ and $U_f$ showing the metallic and insulating states
as discussed above (Fig.5 for $n_{f1} = n_{f2} = 1/3$).
In Fig.6 we show the corresponding phase diagram in $U$-$U_f$ plane
for $n_{f1}=n_{f2}=1/3$.

\begin{figure}
\begin{center}
\vspace{-0.40cm}
\includegraphics[width=6.80cm, height=5.80cm]{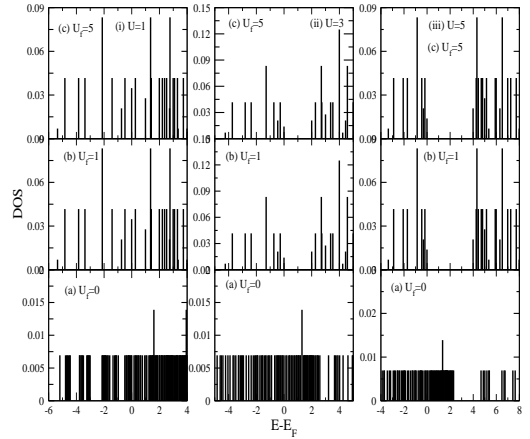}
\caption{ Density of states for different value of $U$ and $U_f$ at $n_{f_{1}} = n_{f_{2}} = 1/3$.}
\end{center}
\end{figure}

\begin{figure}
\begin{center}
\includegraphics[width=7.0cm, height=6.0cm]{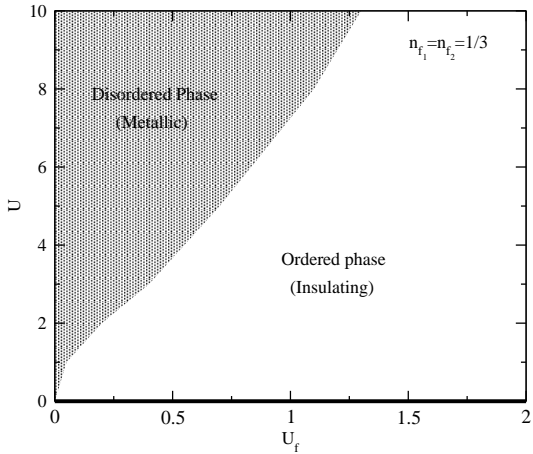}
\caption{Phase diagram showing metallic and insulating regions for $n_{f_{1}}$ = $n_{f_{2}}$
= 1/3, in $U-U_f$ plane. The $x$-axis ($U=0$ line) is trivially metallic.}
\end{center}
\end{figure}

\begin{figure}
\begin{center}
\includegraphics[width=7.0cm,height=6.0cm]{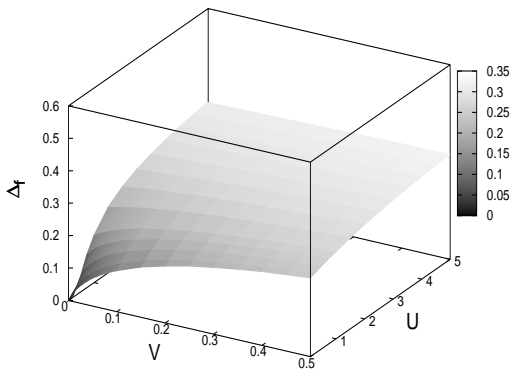}
\caption{\label{fig:wide} The mean-field excitonic order
parameter ${\Delta}_f$ in the $V$-$U$ plane for $n_{f1}=n_{f2}=1/3$.}
\end{center}
\end{figure}

\begin{figure}
\begin{center}
\includegraphics[width=6.0cm]{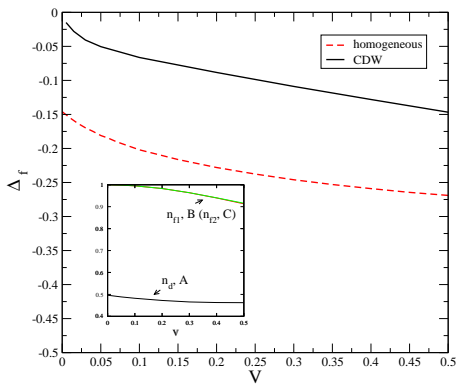}
\caption{\label{fig:wide} (colour online) The mean-field excitonic order parameter for $n_{f1}=n_{f2}=1/3$ and $U=1$
in the homogeneous (lower curve) and ordered (upper curve) ground states. Inset shows
the orbital densities at A, B and C sublattice respectively.}
\end{center}
\end{figure}

The presence of the local $U(1)$ gauge symmetry for each of the two
localized levels have additional advantages. As the conservation of
$f^{\dagger}_{i\alpha}f_{i\alpha}$ renders the $d-$electron part
diagonalisable (in a `local' potential of the $f-$electrons), the problem
then becomes exactly solvable in the infinite dimensional limit where
the $d-$electron self-energy is local~\cite{metz}. Such a solution has been
obtained
by Brandt and Mielsch~\cite{brandt2} for the original Falicov-Kimball model.
A straightforward generalization is possible in the present model as well.
The local Green's function, in the limit of infinite spatial dimension,
turns out to be
\begin{eqnarray}
G(\omega_{n})=<(1-n_{f1})(1-n_{f2})>G_{0}(\omega_{n})
\nonumber \\  
+ <n_{f1}(1-n_{f2})
+n_{f2}(1-n_{f1})>(G^{-1}_{0}(\omega_{n})-U)^{-1}
\nonumber \\
+ <n_{f1}n_{f2}>(G^{-1}_{0}(\omega_{n})-2U)^{-1}.
\end{eqnarray}
\noindent where $G_{0}(\omega_{n})=(i\omega_{n}+\mu+\delta(\omega_{n}))^{-1}$
with $\omega_{n}=(2n+1)\pi KT$ the Fermionic Matsubara
frequency, $\delta(\omega_n)$ the self-consistent, time-dependent
generalized potential.
Inserting the Dyson equation $G^{-1}_{0}(\omega_{n})=G^{-1}(\omega_{n})+
\Sigma(\omega_{n})$ in the above, a cubic equation for the self-energy
$\Sigma(\omega_{n})$ can be written. If one further takes the limit of
$U_{f}\rightarrow \infty$, then the last term on the right side in Eqn.(3) above vanishes
and writing $n_{f}=n_{f1}+n_{f2}$, one gets back the Brandt-Mielsch solution.
Note that the Green's function appears identical to that of the coherent
potential approximation~\cite{velic} (as also alloy analogy, Hubbard-III),
but the difference here is that $<n_{f\alpha}>$ are determined
by the Green's function $G(\omega_{n})$ through $n_{f}=f(\tilde {E}_{f\alpha})$
and $\tilde{E}_{f\alpha}= E_{f\alpha}-KT \sum_{\omega_{n}}\log (1-UG_{0}(\omega_{n}))$
(here $f(x)$ is the Fermi function $[1+exp(x-\mu)/KT]^{-1}$).
In this limit, the self-energy is easily obtained in the case where the
filling is $<n_{f1}>=<nf_{2}>=<n_{d}>=\frac{1}{3}.$ The choice of Hartree
self-energy $\Sigma=U<n_{f}>$ evidently minimizes the ground state energy.
The corresponding configuration is an ordered three-sublattice
arrangement of $d,\,f_{1}$ and $f_{2}$ electrons (with $d,\,f_{1},\,f_{2}$ electrons
occupying A, B, C sublattices), in the limit
U$_{f}\rightarrow \infty$ where the two types of $f-$electrons must
not occupy the same site. In fact the argument,
following Czycholl~\cite{czych}, can be extended (for the ground state
at least) to the case where a small hybridization term like $Vd^{\dagger}_{i}
f_{i\alpha}$ breaks the local invariance of $f_{\alpha}-$electron number,
though the global U(1) symmetry is still extant. We turn to this situation now.

There have been several studies on the FKM on a square lattice over the years
with a hybridization term that mixes localized and extended electrons. An
important issue in this context was raised by Portengen et al.,~\cite{port}
regarding the spontaneous symmetry breaking (SSB) in such a model. A term
$Vd^{\dagger}_{i}f_{i,\alpha}$ would induce non-zero excitonic (or ferroelectric
~\cite{port}) averages like $<d^{\dagger}_{i}f_{i,\alpha}>$ as long
as V is non zero. It was shown~\cite{port} from a mean-field theory calculation on a
square lattice at half-filling that this average tends to a finite value even
when $V \rightarrow 0$, leading to a spontaneously broken symmetry in the
ground state~\cite{comm}.
Although the soft local dynamic fluctuations between $f_{\alpha}$ and $d$ electrons in
the limit $V \rightarrow 0$ cannot be treated properly in the static mean-field,
we nevertheless undertake a similar mean-field analysis on the triangular lattice
following previous work~\cite{czych,port}. Assuming a
homogeneous ground state and excitonic mean-fields, we find (Fig.7) that when
$U$ is finite, averages of the type $\Delta_{f,\alpha} = <d^{\dagger}_{i} f_{i\alpha}>$ do
not vanish in the limit $V\rightarrow 0$, indicating an SSB  similar to the
situation on a square lattice~\cite{port}. Such a non-zero average would, then,
imply a homogeneously mixed valent ground state.

However, the assumption of a homogeneous ground state itself is called into
question for the ranges of parameters being studied. It is
important to note that there could possibly be other ordered states with lower
energy at half-filling and one would therefore need to first ascertain the correct
ground state. Czycholl~\cite{czych} had shown that on the square lattice, this is
indeed the case and the right ground state is an ordered
two-sublattice charge density state. He also showed that such a ground state does
not support SSB and a homogeneous mixed valence.
It is not clear a priori that on a triangular lattice, similar considerations would
apply. Clearly our mean-field analysis above shows that on a homogeneous ground state,
the results of Portengen et al. hold. We, therefore, look for the ground state
in the mean-field theory and work out the possibility of an SSB on a triangular
lattice.

Following the arguments in the infinite dimensional limit above, we expect an ordered
three-sublattice orbital density state to appear at half-filling when
$n_{d}=n_{f,\alpha}=1/3$ even for small finite $V$. We set up the mean-field calculation
to look for an orbitally ordered state and allow for inhomogeneous local order parameters.
This involves calculating all the local order parameters in the real space iteratively
for self-consistency. We describe the result in Fig.8 where an inhomogeneous mean-field
solution with a three-sublattice orbital structure emerges as the lowest energy state.
The orbital density order parameter is shown in the
inset of Fig.8. The A sublattice has negligible $f$-occupancy and a finite $d$-electron
density, while B (C) sublattice has predominantly $f_1$ \, ($f_2$) occupation and
a small $d$-occupation. Clearly as $V$ rises, such a state becomes less stable and at a
critical $V$ the orbital order would melt (although mean-field analysis may not be quite
valid in that range) due to strong quantum fluctuations. On reducing $V$, we observe that in
the limit of $V\rightarrow 0$, the excitonic order parameter (note $\Delta_{f,\alpha} =
<d^{\dagger}_{i} f_{i\alpha}>$ is same for $\alpha=1,2$ in the degenerate limit
considered) vanishes leading to an absence of SSB in this case. For comparison, we
also show the behavior of this order parameter in a homogeneous ground state. In this
case the solution supports a non-vanishing expectation value even at $V\rightarrow 0$.

In conclusion, we have studied an extended Falikov-Kimball model with two
localized states on a triangular lattice. Such a model is found to reproduce
orbitally ordered insulating ground states as reported (or predicted) in experimental
systems like GdI$_2$~\cite{simon}, NaTiO$_2$ etc. Furthermore, the disorder to order transitions
have also been seen to accompany a metal-insulator transition. Many of these orbitally
ordered ground states would induce spontaneous lattice distortions due to the
breaking of inversion symmetry and lead to ferroelectricity driven not by the conventional
mechanisms, but by electronic correlations instead. We also investigate the possibility of 
a spontaneously broken symmetry in the form of an excitonic order paramete in this Hamiltonian 
and resolve that there is no SSB in the ground state on a triangular lattice.

\acknowledgments  UKY acknowledges CSIR, India for a research fellowship. AT thanks
M. Laad for useful discussions.

\end{document}